\documentclass[prd,twocolumn,twoside,preprintnumbers,superscriptaddress,nofootinbib]{revtex4}

\pdfoutput=1
\usepackage{amsmath,amssymb,bm,graphicx,bbold,epsf,colordvi}
\usepackage{lipsum}
\allowdisplaybreaks 
\usepackage{color}
\newcommand\f{\frac}
\newcommand{\ba}{\begin{eqnarray}}
\newcommand{\ea}{\end{eqnarray}}
\newcommand{\be}{\begin{equation}}
\newcommand{\ee}{\end{equation}}
\newcommand{\nn}{\nonumber}

\def\d{{\rm d}}
\def\OMIT#1{{}}

\newcommand{\mcdot}{\!\cdot\!}

\newcommand{\bra}[1]{\left\langle #1\right\rvert}
\newcommand{\ket}[1]{\left\lvert #1\right\rangle}

\newcommand{\e}{\mathrm{e}}

\newcommand{\vc}[1]{\boldsymbol{#1}}

\def\SCETG{${\rm SCET}_{\rm G}\,$}

\begin{document}
%%%%%%%%%%%%%%%%%%%%%%%%%%%%%%%%%

\preprint{\vbox{\hbox{ACFI-T15-18}}}

\title{Initial-state splitting kernels in cold nuclear matter}

\author{Grigory Ovanesyan}
\affiliation{Physics Department, University of Massachusetts Amherst, Amherst, MA 01003, USA}%
\author{Felix Ringer}
\affiliation{Theoretical Division, MS B283, Los Alamos National Laboratory, Los Alamos, NM 87545, USA}%
\author{Ivan Vitev}
\affiliation{Theoretical Division, MS B283, Los Alamos National Laboratory, Los Alamos, NM 87545, USA}%

%%%%%%%%%%%%%%%%%%%%%%%%%%%%%%%%%%
%%%%%%%%%%%%%%%%%%%%%%%%%%%%%%%%%%
\begin{abstract}
We derive medium-induced splitting kernels for energetic partons that undergo interactions in dense QCD matter before a hard-scattering event at large momentum transfer $Q^2$. Working in the framework of the effective theory \SCETG, we compute the splitting kernels beyond the soft gluon approximation. We present numerical studies that compare our new results with previous findings. We expect the full medium-induced splitting kernels to be most relevant for the extension of initial-state cold nuclear matter energy loss phenomenology in both p+A and A+A collisions. 
\end{abstract}

\maketitle
\section{Introduction}\label{sec:introduction}

Initial-state cold nuclear matter (CNM) effects in A+A and in particular in p+A collisions have received growing attention in recent years~\cite{Albacete:2013ei,Andronic:2015wma}. Various CNM effects have been discussed extensively in the literature~\cite{Gavin:1991qk,Eskola:1998df,Vogt:1999dw,Johnson:2000ph,Wang:2001ifa,Gyulassy:2002yv,Accardi:2002ik,Qiu:2004qk,Qiu:2004da,Kopeliovich:2005ym,Hirai:2007sx,Vitev:2007ve,Neufeld:2010dz,Xing:2011fb,Arleo:2012hn,Kang:2012kc,Kang:2013ufa,Kang:2014hha,Kang:2015mta}. 
Coherent and incoherent scattering, which lead to power corrections in the production cross sections and the Cronin effect, are limited to small and moderate transverse momenta $p_T$. Nuclear shadowing effects and inelastic multi-parton processes in a QCD medium can lead to modification of the production cross sections for particles and jets at large transverse momenta. In particular, the effect of CNM energy loss can be amplified at high mass or $p_T$ near the kinematic bounds~\cite{Neufeld:2010dz,Kang:2015mta}.
Here, energy loss refers to the medium-induced soft bremsstrahlung processes that redistribute part of the energy of the leading parton in a shower of soft gluons. Generally speaking, this can happen before or after the hard-scattering. The beam jets of the incoming nucleon can also lose energy via Bertsch-Gunion bremsstrahlung in a process that contributes to the generation of the underlying event multiplicity for the hadronic collisions~\cite{Gyulassy:2014cfa}. 

Renewed interest in CNM effects was sparked by recent experimental results that revealed a large and highly non-trivial nuclear modification of jet production yields in p+Pb collisions at the LHC~\cite{Adam:2015hoa,ATLAS:2014cpa}, as well as in d+Au at RHIC~\cite{Perepelitsa:2013jua}. These nuclear modifications manifested themselves in a suppression of the jet production cross section for central collisions, whereas the experiments found an enhancement for peripheral collisions. While it is not yet clear how much of these effects arise from centrality bias, CNM energy loss can contribute to the jet suppression in central and semi-central collisions. In particular, the scaling behavior of the observed suppression as a function of $p_T\cosh(y)$ can be understood in this picture, where $y$ is the rapidity of the observed jet. CNM energy loss may also be a non-negligible effect in describing nuclear modifications in A+A collisions, even though it is suppressed compared to final-state energy loss. 
In the framework of the reaction operator approach~\cite{Gyulassy:2002yv}, CNM energy loss was first computed in~\cite{Vitev:2007ve} and applied to Drell-Yan production \cite{Neufeld:2010dz}. These initial-state energy loss calculations are an extension of the Gyulassy-Levai-Vitev approach to energy loss for the final state, which successfully predicted and described the magnitude, centrality and energy dependence of jet quenching in A+A collisions, e.g. see~\cite{Vitev:2004gn,Dai:2012am}. It is certainly of great relevance to theory and experiment to further improve the framework of inelastic parton scattering in nuclear matter, which we address in this work.

Advances in understanding inelastic parton processes in dense QCD matter are enabled by Soft Collinear Effective Theory~\cite{Bauer:2000ew, Bauer:2000yr, Bauer:2001ct, Bauer:2001yt,Beneke:2002ph} (SCET) and its extension to accommodate jet propagation in a nuclear medium. In recent years SCET has become a valuable tool for describing energetic particle and jet production at present-day collider experiments~\cite{Fleming:2007xt, Becher:2008cf, Hornig:2009vb, Abbate:2010xh, Becher:2007ty, Stewart:2009yx, Becher:2009th, Chien:2015cka}. An effective theory describing the interaction of collinear partons with a QCD medium, Soft Collinear Effective Theory with Glauber gluons (\SCETG), was developed in~\cite{Idilbi:2008vm,Ovanesyan:2011xy} and theoretical results for the elastic and inelastic soft parton scattering processes given in~\cite{DEramo:2010ak,Ovanesyan:2011kn,Fickinger:2013xwa}. The basic idea is to include a ``Glauber mode'' in the effective theory that describes the interaction of energetic partons in matter only by a transverse momentum exchange. In the traditional approach to energy loss, medium-induced splittings were derived only in the small-$x$, or soft gluon limit. Here, $x$ denotes the large lightcone momentum fraction carried away by the emitted parton and $x \ll 1$ denotes the approximation in which non-Abelian energy loss is well defined and discussed in the literature. \SCETG allows to systematically go beyond this approximation to finite values of $x$ and understand in detail the formation of an in-medium parton shower. In~\cite{Ovanesyan:2011xy,Ovanesyan:2011kn,Fickinger:2013xwa} the full medium-induced splitting kernels were derived for final-state 
parton showers. They describe the branching processes that follow a large $Q^2$ hard-scattering and result in energetic particles and jets seen in the detectors. The obtained full splitting kernels have been successfully applied in order to improve the theoretical precision of the computation of various observables in A+A collisions, including jet and hadron production cross sections~\cite{Kang:2014xsa,Chien:2015vja} as well as jet cross sections and shapes~\cite{Chien:2014nsa,Chien:2015hda}. In the first two papers listed here, an in-medium DGLAP QCD evolution approach was used, in analogy to the evolution of fragmentation functions in the vacuum. In this work, we derive the analogous finite-$x$ medium-induced splitting kernels for the initial state, relevant to parton shower formation in cold nuclear matter. In the future, we plan to extend the existing CNM energy loss phenomenology using DGLAP evolution techniques for the parton distribution functions. 

The remainder of this paper is organized as follows. In Section~\ref{sec:framework}, we recall the basic framework of \SCETG and rederive the vacuum splitting kernels. In Section~\ref{sec:initialstate}, we present our results for the full in-medium splitting kernels in the initial state, as well as their soft-gluon emission limit. In particular, we point out important differences to the final state results. We present numerical studies comparing the new splitting kernels with their small-$x$ approximations in Section~\ref{sec:numerics}. Finally, we draw our conclusions in Section~\ref{sec:conclusions}.

\section{Theoretical Framework} \label{sec:framework}

The effective theory \SCETG as derived in~\cite{Ovanesyan:2011xy} is based on the following Lagrangian
\ba
\mathcal{L}_{\text{\SCETG}}(\xi_n, A_n, A_G)\,&=&\,\mathcal{L}_{\text{SCET}}(\xi_n, A_n) \nn \\
&&+\mathcal{L}_{\text{G}}\left(\xi_n, A_n, A_G\right) \,,
\ea
where the first term is the usual SCET Lagrangian~\cite{Bauer:2000ew, Bauer:2000yr, Bauer:2001ct, Bauer:2001yt} and the second term describes the physical interaction of collinear partons with the QCD medium via the exchange of Glauber gluons. The explicit form of the Lagrangians as well as the corresponding Feynman rules can be found in~\cite{Ovanesyan:2011xy}. In this work, we present results for the initial-state splitting processes $q\to qg$, $q\to gq$, $g\to gg$ and $g\to q\bar q$. Their respective amplitudes are given by
\ba
A_{q\rightarrow ab}\, &=&\, \bra{a(p) b(k)}T\,\e^{iS}\,\bar{\chi}_n(x_0)\ket{q(p_0)}, \nn \\
A_{g\rightarrow ab}\, &=&\, \bra{a(p) b(k)}T\,\e^{iS}\,\mathcal{B}^{\lambda c}_n(x_0)\ket{g(p_0)},\label{Adef}
\ea
where $\chi, \mathcal{B}$ are collinear gauge invariant SCET fields for quarks and gluons respectively~\cite{Arnesen:2005nk, Bauer:2008qu} and $S$ is the \SCETG action. The partons after the splitting are labeled as $a,b$, which is representative for the two possible splittings of a quark or a gluon as listed above. The momenta of the involved partons are given in parentheses in Eq.~(\ref{Adef}) and are related by momentum conservation $p_0=p+k$. The parton with momentum $k$ carries away a fraction $x=k^+/p_0^+$ of the energy of the initial parton with momentum $p_0$. In this work, we are considering massless partons only. Using the on-shell conditions $p_0^2=k^2=0$, we may parametrize the involved momenta as
\begin{eqnarray}\label{eq:pkp0}
&&p_0 \, = \, \left[p_0^+, 0,\vc 0_\perp \right], \nn \\
&&k \,=\, \left[x p_0^+, \f{\vc k_\perp^2}{xp_0^+},\vc k_\perp\right],\nn  \\
&&p \,=\, p_0-k = \left[(1-x)p_0^+, \f{-\vc k_\perp^2}{ xp_0^+},-\vc k_\perp\right] ,
\end{eqnarray}
where we have adopted the following notation for light-cone four-vectors $q=[q^+,q^-,\vc q_\perp]=[\bar n\cdot q,n\cdot q,\vc q_\perp]$ for any vector $q$ and $n^{\mu}=\left(1,0,0,1\right)$, $\bar{n}^{\mu}=\left(1,0,0,-1\right)$. Note that in Eqs.~(\ref{eq:pkp0}), we have chosen the positive light-cone direction along the momentum direction of the parent parton.

Using only the SCET Lagrangian, we can reproduce again the results for the vacuum leading-order Altarelli-Parisi splitting kernels~\cite{Altarelli:1977zs}. As in~\cite{Ovanesyan:2011kn}, we obtain
\begin{eqnarray}
\label{qqg}
   \left(\frac{\d N}{\d x \,\d^2 \vc k}_{\perp}\right)_{q\rightarrow qg}&=&
\frac{\alpha_s}{2\pi^2} C_F \frac{1+(1-x)^2}{x}\frac{1}{{\vc k}_{\perp}^2},\\
\label{ggg}
    \left(\frac{\d N}{\d x \,\d^2{\vc k}}_{\perp}\right)_{g\rightarrow gg}&=& \frac{\alpha_s}{2\pi^2}
  2 C_A \Big(\frac{1-x}{x}+\frac{x}{1-x}  \nonumber\\
&&   \qquad\qquad +x(1-x) \Big)\frac{1}{{\vc k}_{\perp}^2},\\
\label{gqq}
    \left(\frac{\d N}{\d x \,\d^2{\vc k}}_{\perp}\right)_{g\rightarrow q\bar{q}}&=
&  \frac{\alpha_s}{2\pi^2} T_R\  \left( x^2+(1-x)^2 \right)\frac{1}{{\vc k}_{\perp}^2},\\
   \left(\frac{\d N}{\d x \,\d^2{\vc k}}_{\perp}\right)_{q\rightarrow gq}&=&
\label{qgq}
  \left(\frac{\d N}{\d x \,\d^2{\vc k}}_{\perp}\right)_{q\rightarrow qg} (x\rightarrow 1-x).\nonumber\\
 \end{eqnarray}
In this work, we are only considering real splitting processes away from the endpoints at $x=1$ and $x=0$. However, in future phenomenological applications, we are going to employ generalized DGLAP evolution techniques using the in-medium initial-state splitting kernels. An analogous technique for the final-state splitting kernels was used in~\cite{Kang:2014xsa,Chien:2015vja} for studies of jet quenching phenomenology in A+A collisions. In this context, the in-medium virtual contributions were derived from momentum and flavor sum rules satisfied by the splitting kernels. The same techniques can be used for the initial-state splitting kernels.

\section{Initial-state parton splittings} \label{sec:initialstate}

\begin{figure}[!t]
\includegraphics[width=230pt]{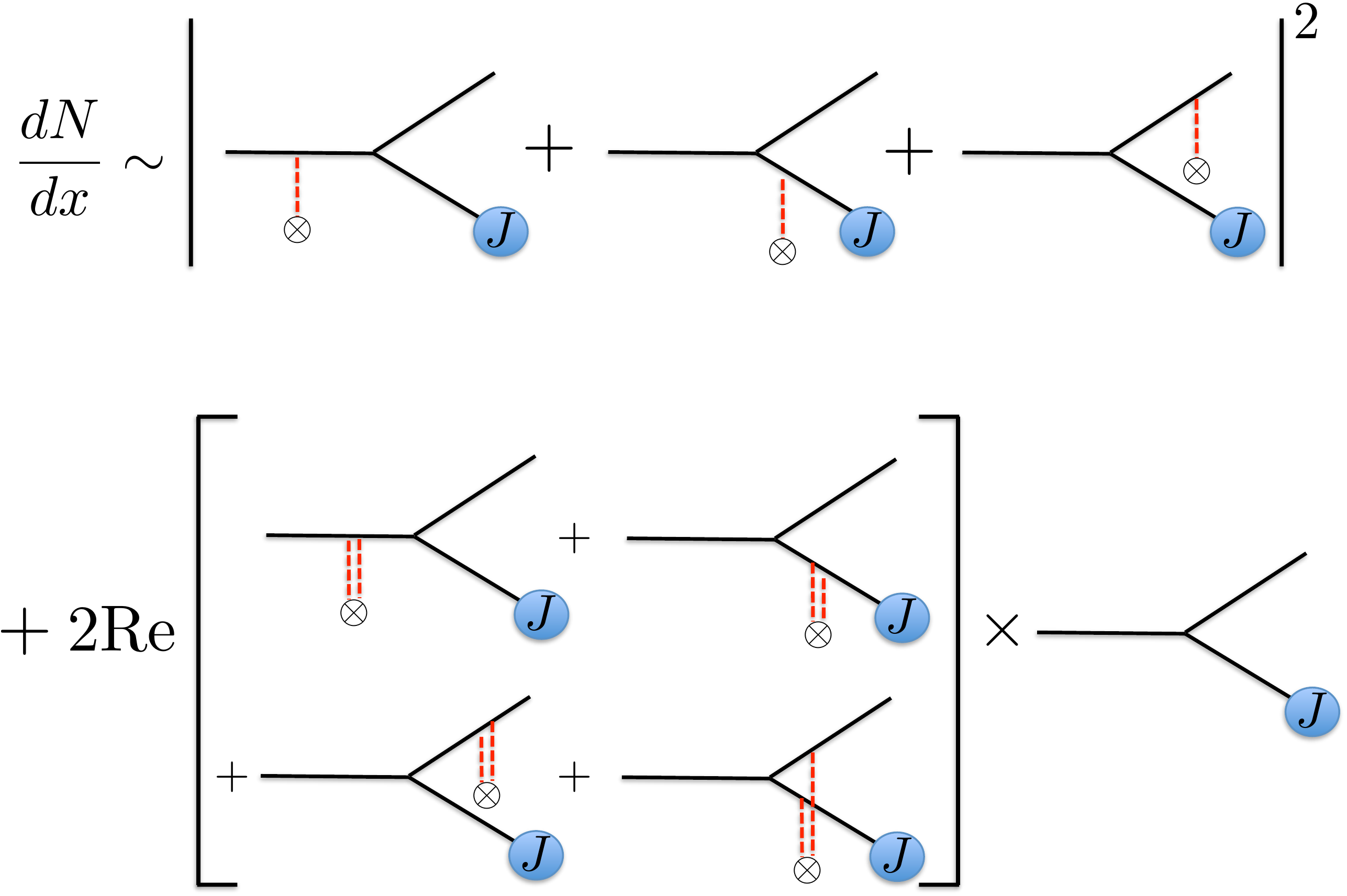}
\caption{Feynman diagrams contributing to the initial-state medium-induced splittings kernels at first order in opacity. Red lines correspond to the exchange of Glauber gluons with the medium. The kinematics and topology are common to all splitting processes: $q\rightarrow qg$, $q\rightarrow gq$, $g\rightarrow gg$, $g\rightarrow q \bar{q}$.}
\label{fig:diagrams}
\end{figure}

In Ref.~\cite{Ovanesyan:2011xy}, it was shown that a particularly simple gauge choice for calculations of jets in the medium is the so-called hybrid gauge. In this gauge the collinear gluons are treated in the light-cone gauge, while the Glauber gluons are treated in the covariant gauge. With this choice both the collinear Wilson line as well as the transverse gauge link which appear in the effective theory reduce to unity. This choice simplifies our calculations considerably. In order to compute the splitting kernels to first order in opacity, we need to consider single- and double-Born diagrams. The topology of the relevant diagrams for initial-state splitting processes is shown in Fig.~\ref{fig:diagrams}. In the first line of Fig.~\ref{fig:diagrams}, the single-Born diagrams are shown, which have to be squared. In the second row, the non-zero double-Born diagrams are shown which appear as interference terms with the vacuum Born amplitude.

We now present our results for the initial-state in-medium splitting kernels, which constitutes the main result of this paper. In order to simplify our notation, we define the following two dimensional transverse vectors $\vc A_\perp,\;\vc B_\perp,\;\vc C_\perp$ and phases $\Omega_{1,2,3}$ similar to the notation in~\cite{Ovanesyan:2011xy,Ovanesyan:2011kn}
\begin{eqnarray}
&&\vc{A}_{\perp}={\vc k}_{\perp},\;\; \vc{B}_{\perp}={\vc k}_{\perp}-x{\vc q}_{\perp},\nonumber\\
&&\vc{C}_{\perp}={\vc k}_{\perp}-{\vc q}_{\perp},\nonumber\\
&&\Omega_1-\Omega_2=\frac{\vc{B}^2_{\perp}}{x(1-x)p^+}, \;\; \Omega_3=\frac{\vc{A}^2_{\perp}}{x(1-x)p^+}\,,
\end{eqnarray}
where $\vc q_\perp$ is the transverse momentum of the partons obtained due to Glauber gluon exchange with the medium. For the two splitting processes $q\to qg$ and $q\to gq$, we obtain
\begin{widetext}
\begin{eqnarray}\label{eq:qqg}
&&\left(\frac{\d N}{\d x\,\d^2{\vc k}_{\perp}}\right)_{q\rightarrow qg}\nonumber\\
&&\qquad\qquad=C_F\frac{\alpha_s}{2\pi^2} \frac{1+(1-x)^2}{x}\int\frac{\d z}{\lambda_{g}(z)}\,\d^2{\vc q}_{\perp}\frac{1}{\sigma_{\text{el}}}\frac{\d\sigma_{\text{el}}}{\d^2{\vc q}_{\perp}}\Bigg[\frac{\vc{B}_{\perp}}{\vc{B}_{\perp}^2}\mcdot \left(\frac{\vc{B}_{\perp}}{\vc{B}_{\perp}^2}-\frac{\vc{C}_{\perp}}{\vc{C}_{\perp}^2}\right)\left(1-\cos(\Omega_1-\Omega_2)\Delta z\right) \nonumber\\
&&\qquad\qquad+\f{1}{\vc C_\perp^2}-\f{1}{\vc A_\perp^2}+\f{\vc A_\perp}{\vc A_\perp^2}\mcdot\left(\f{\vc A_\perp}{\vc A_\perp^2}-\f{\vc C_\perp}{\vc C_\perp^2}\right)\left(1-\cos\Omega_3\Delta z\right)-\frac{1}{N_c^2}\f{\vc B_\perp}{\vc B_\perp^2}\mcdot \left(\f{\vc B_\perp}{\vc B_\perp^2}-\f{\vc A_\perp}{\vc A_\perp^2}\right) \left(1-\cos(\Omega_1-\Omega_2)\Delta z\right)\Bigg] \, , \\
&&\left(\frac{\d N}{\d x\,\d^2{\vc k}_{\perp}}\right)_{q\rightarrow gq}\nonumber\\
&&\qquad\qquad=C_F\frac{\alpha_s}{2\pi^2}\frac{1+x^2}{1-x}\int\frac{\d z}{\lambda_{q}(z)}\,\d^2{\vc q}_{\perp}\frac{1}{\sigma_{\text{el}}}\frac{\d\sigma_{\text{el}}}{\d^2{\vc q}_{\perp}}\Bigg[2\frac{\vc{B}_{\perp}}{\vc{B}_{\perp}^2}\mcdot \left(\frac{\vc{B}_{\perp}}{\vc{B}_{\perp}^2}-\frac{\vc{C}_{\perp}}{\vc{C}_{\perp}^2}\right)\left(1-\cos(\Omega_1-\Omega_2)\Delta z\right)+\frac{1}{{\vc C}_{\perp}^2}-\frac{1}{{\vc A}_{\perp}^2}\nonumber\\
&&\qquad\qquad+\frac{2N_c^2}{N_c^2-1}\left(\frac{\vc{A}_{\perp}}{\vc{A}_{\perp}^2}-\frac{\vc{C}_{\perp}}{\vc{C}_{\perp}^2}\right)\mcdot \left(\frac{\vc{A}_{\perp}}{\vc{A}_{\perp}^2}\left(1-\cos\Omega_3\Delta z\right)-\frac{\vc{B}_{\perp}}{\vc{B}_{\perp}^2}(1-\cos\left(\Omega_1-\Omega_2\right)\Delta z)\right)\Bigg] \, .
\end{eqnarray}
\end{widetext}
Here, $\lambda_{g,q}(z)$ is the scattering length of a gluon or quark in the medium respectively, which are related by $\lambda_q/\lambda_g=C_A/C_F$ to lowest order. Furthermore, we need to take the integral over the interaction region $0<\Delta z<L$, where $L$ is the size of the QCD medium and $(1/\sigma_{\text{el}})\,\d\sigma_{\text{el}}/\d^2{\vc q}_{\perp}$ is the normalized in-medium elastic scattering cross section. Note that independent of the parton type, the elastic cross section is dominated by $t$-channel gluon exchange at high energies. 

Similar to the final-state results, the two splitting kernels for the processes $g\to gg$ and $g\to q\bar q$ differ only by their respective color factors and the overall vacuum splitting kernel
\begin{widetext}
\begin{eqnarray}\label{eq:ggggqq}
&&  \left( \frac{\d N}{ \d x\,\d^2\vc{k}_{\perp} }\right)_{ \left\{ \begin{array}{c}   g \rightarrow gg\\     g\rightarrow q\bar{q}  \end{array} \right\} } 
= 
 \left\{ \begin{array}{c}     \frac{\alpha_s}{2\pi^2} \, 2 C_A \left(\frac{x}{1-x}+\frac{1-x}{x}+x(1-x) \right)  \\[1ex]
                           \frac{\alpha_s}{2\pi^2}  T_R \left( x^2+(1-x)^2 \right)  \end{array} \right\}
\int {\d\Delta z}   \left\{ \begin{array}{c}    \frac{1}{\lambda_g(z)} \\[1ex]  \frac{1}{\lambda_q(z)}    \end{array} \right\}
\int \d^2 \vc q_\perp  \frac{1}{\sigma_{el}} \frac{\d\sigma_{el}^{\; {\rm medium}}}{\d^2 \vc q_\perp} \;  \nonumber \\
&& \qquad \qquad
\times \Bigg[ \left(2\f{\vc B_\perp}{\vc B_\perp^2}-\f{\vc A_\perp}{\vc A_\perp^2}-\f{\vc C_\perp}{\vc C_\perp^2}\right)^2 +2 \left(\f{\vc A_\perp}{\vc A_\perp^2}+\f{\vc C_\perp}{\vc C_\perp^2}\right)\mcdot\left(\f{\vc B_\perp}{\vc B_\perp^2}-\f{\vc A_\perp}{\vc A_\perp^2}\right) - 2 \f{\vc B_\perp}{\vc B_\perp^2}\mcdot \left(2\f{\vc B_\perp}{\vc B_\perp^2}-\f{\vc A_\perp}{\vc A_\perp^2}-\f{\vc C_\perp}{\vc C_\perp^2} \right) \nn \\
 &&\qquad \qquad \times \cos[(\Omega_1-\Omega_2)\Delta z] + \left\{ \begin{array}{c}   - \frac{1}{2}   \\[1ex] \frac{1}{N_c^2-1} \end{array} \right\}
\Bigg(2 \f{\vc B_\perp}{\vc B_\perp^2}\mcdot\left(2\f{\vc B_\perp}{\vc B_\perp^2}-\f{\vc A_\perp}{\vc A_\perp^2}-\f{\vc C_\perp}{\vc C_\perp^2}\right)\left(1-\cos[(\Omega_1-\Omega_2)\Delta z] \right)\nn\\
 &&\qquad \qquad  - 2 \f{\vc A_\perp}{\vc A_\perp^2}\mcdot\left(\f{\vc A_\perp}{\vc A_\perp^2}-\f{\vc C_\perp}{\vc C_\perp^2}\right) \left(1-\cos[\Omega_3\Delta z]\right) \Bigg) \Bigg] \, .      
\label{CohRadSX2} 
\end{eqnarray} 
\end{widetext}
All four initial-state splitting kernels presented here are proportional to their vacuum counter parts. This feature is shared with the final-state in-medium splitting kernels. Note that in the vacuum, we can obtain the splitting kernel $q\to gq$ from $q\to qg$ by substituting $x\to 1-x$, cf. Eq.~(\ref{qgq}). In~\cite{Ovanesyan:2011kn}, it was found that also the final-state in-medium splitting kernels possess this symmetry. In addition, both $g\to gg$ and $g\to q\bar q$ are symmetric under $x\to 1-x$. As it turns out, this is not the case anymore for the initial-state in-medium splitting kernels at the differential level. Our results for the splitting processes $q\to qg$ and $q\to gq$ are not related via $x\to 1-x$ and the results presented here for $g\to gg$ and $g\to q\bar q$ are not invariant under the transformation $x\to 1-x$. This somewhat surprising asymmetry can be traced back to the fact that for an initial-state splitting process one of the partons is off-shell after the splitting. For final-state splitting kernels, both outgoing partons are on-shell. In the medium, we obtain different splitting kernels depending on which final-state parton is off-shell. That being said, we do find that in the case of the infinite limits of the integrations in the variables $\vc{k}_{\perp}, \vc{q}_{\perp}$, the fully inclusive initial-state splitting kernels do obey the same $x\rightarrow 1-x$ symmetries as in the vacuum and for the final-state splitting kernels. To see this analytically, consider the transformation\footnote{In addition to this transformation it is essential that the following integral vanishes exactly with the infinite limits of integration: $\int \d^2 \vc q_\perp \d^2 \vc k_\perp  \frac{1}{\sigma_{el}} \frac{\d\sigma_{el}^{\; {\rm medium}}}{\d^2 \vc q_\perp}\left(\frac{1}{\vc{C}_{\perp}^2}-\frac{1}{\vc{A}_{\perp}^2}\right)=0$.} $x\rightarrow 1-x, \vc{q}_{\perp}\rightarrow -\vc{q}_{\perp}, \vc{k}_{\perp}\rightarrow \vc{k}_{\perp}-\vc{q}_{\perp}$, which effectively is equivalent to $\vc{A}_{\perp}\leftrightarrow \vc{C}_{\perp}$ and $\vc{B}_{\perp}\rightarrow \vc{B}_{\perp}$. The finite phase space cuts break this symmetry for the inclusive initial-state splitting kernels only very close to $x=0$ and $x=1$.
 \begin{figure*}[!t]
\includegraphics[width= 220pt]{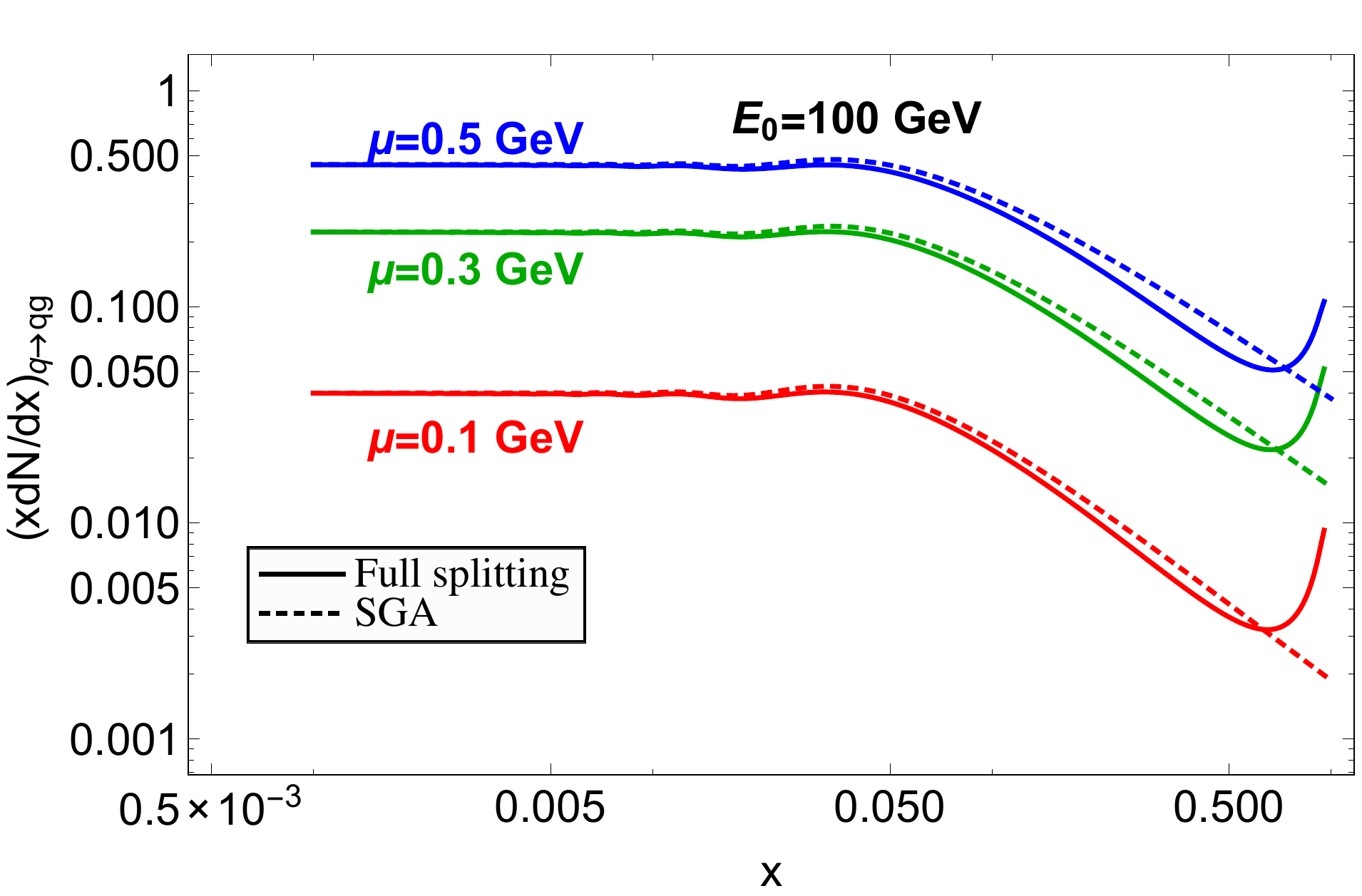} \hspace*{0.1in}\includegraphics[width= 220pt]{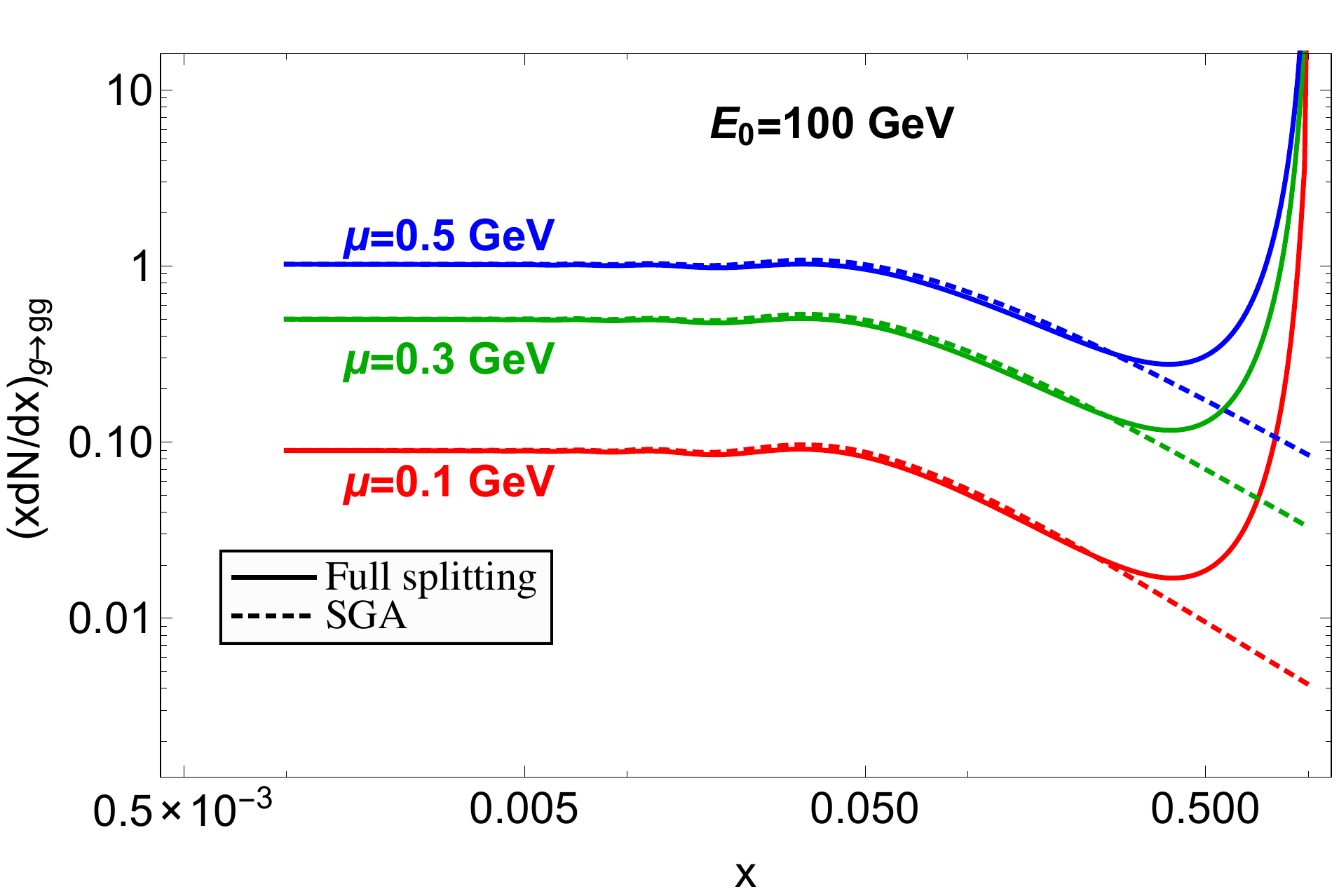}\\
\includegraphics[width= 220pt]{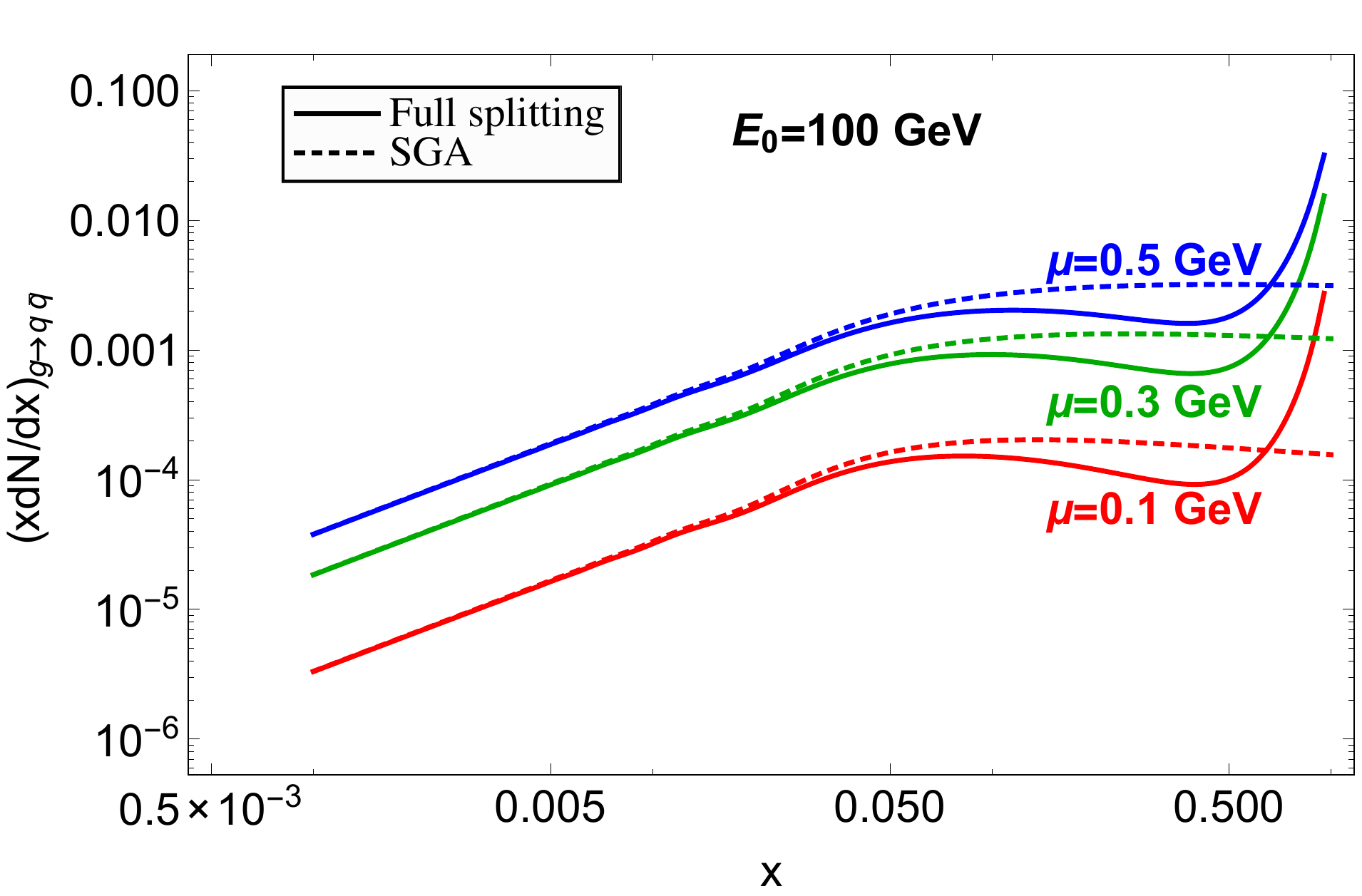} \hspace*{0.1in} \includegraphics[width= 220pt]{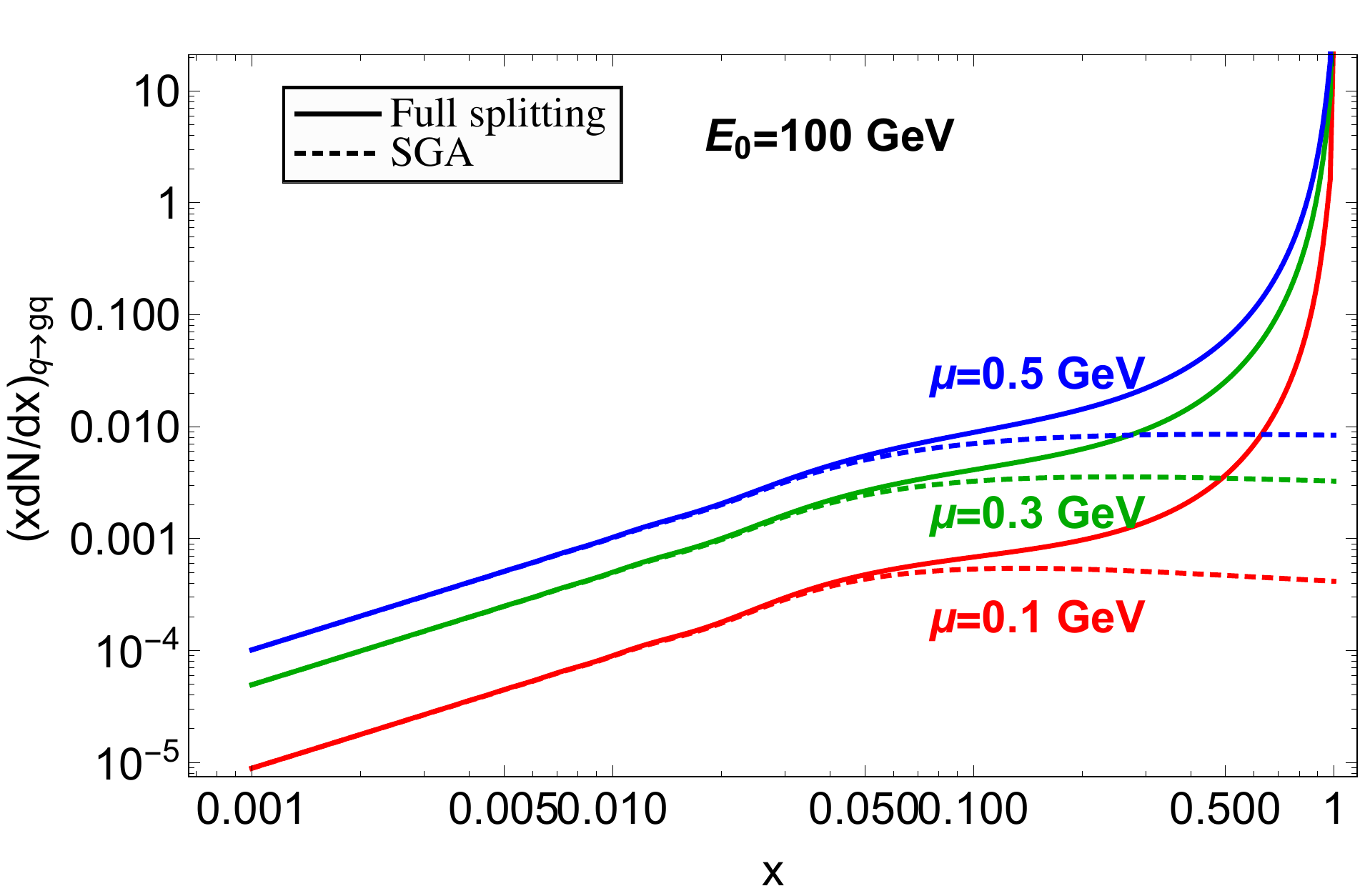}
\caption{The finite-$x$ (solid) and small-$x$ (dashed) intensity spectra $x(\d N/\d x)$ for all four splitting kernels are shown using infinite phase space cuts. We chose a parent parton energy of $E_0=100$~GeV and three different values of the momentum transfer $\mu=0.1,\, 0.3,\, 0.5$~GeV (red, green, blue) using certain assumptions as described in the text.}
\label{fig:numerics1}
\end{figure*}

In order to make the connection with the traditional approach to CNM energy loss of~\cite{Vitev:2007ve}, we take the small-$x$ limit \footnote{Which we will also sometimes call the soft gluon approximation~(SGA).}, $x\ll 1$, of our full results in Eqs.~(\ref{eq:qqg})-(\ref{eq:ggggqq}). Using $\vc A_\perp=\vc B_\perp=\vc k_\perp$, $\vc C_\perp=\vc k_\perp-\vc q_\perp$ and $\Omega_1-\Omega_2=\Omega_3=\vc k^2_\perp/(x p_0^+)$, we find
\begin{eqnarray}
 && \!\!\!\!\!\!\!  x \left( \frac{\d N}{ \d x\,\d^2\vc k_\perp}\right)_{ \left\{ \begin{array}{c}   q \rightarrow qg \\    
 g\rightarrow gg  \\  g \rightarrow q\bar{q} \\ q \rightarrow gq\\   \end{array} \right\} }  =  \frac{\alpha_s}{\pi^2} 
  \left\{ \begin{array}{c}   C_F[ 1+ {\cal O}(x) ] \\  C_A[ 1+ {\cal O}(x) ]  \\
  T_R [0 + \frac{x}{2} + {\cal O}(x^2)]  \\    C_F [0 + \frac{x}{2} + {\cal O}(x^2)]
  \end{array} \right\} \nonumber \\
&&  \;\;  \times    \int {\d\Delta z}   \left\{ \begin{array}{c}    \frac{1}{\lambda_g(z)} \\[1ex]  \frac{1}{\lambda_g(z)}  \\
 \frac{1}{\lambda_q(z)}  \\  \frac{1}{\lambda_q(z)}    \end{array} \right\}
\int \d^2 \vc k_\perp \d^2\vc q_\perp  \frac{1}{\sigma_{el}} \frac{\d\sigma_{el}^{\; {\rm medium}}}{\d^2 \vc q_\perp} \;  \nonumber \\
&&  \;\;  \times \left[\f{\vc q_\perp^2}{\vc k^2_\perp(\vc k_\perp-\vc q_\perp)^2}-2\f{\vc q^2_\perp-\vc k_\perp\mcdot\vc q_\perp}{\vc k^2_\perp(\vc k_\perp-\vc q_\perp)^2} \right]\cos\left( \frac{   \vc k_\perp^2}{xp_0^+} \Delta z\right). \nn�\\
\label{smallx}
\end{eqnarray} 
Note that the two diagonal splitting kernels $q\to qg$ and $g\to gg$ agree with the results in~\cite{Vitev:2007ve}. The off-diagonal splitting kernels for the processes $q\to gq$ and $g\to q\bar q$ are of order ${\cal O}(x)$ and, hence, vanish for $x\to 0$. They are not taken into account in the traditional approach to CNM energy loss. However, we use their ${\cal O}(x)$ expressions for numerical comparisons between the full-$x$ and SGA approximation of the splitting kernels in the next Section~\ref{sec:numerics}. Note that the structure of the overall factors in the curly brackets of~(\ref{smallx}) is identical with the one found for the final-state splitting kernels in~\cite{Ovanesyan:2011kn}.

We would like to point out that partons in the medium acquire an effective mass $m_{\text{eff}}$, which for the final-state splitting kernels is given by thermal masses, while for the initial-state splitting kernels we follow previous treatments and take it to be of the order of the nucleon mass $m_{\text{eff}}=m_N=0.94$~GeV for CNM energy loss~\cite{Vitev:2007ve}. We take this effect into account by substituting $|\vc k_\perp|^2\to |\vc k_\perp|^2+m^2_{\text{eff}}$ for all in-medium splitting kernels for finite-$x$ in Eqs.~(\ref{eq:qqg})-(\ref{eq:ggggqq}) as well as for the small-$x$ results in Eq.~(\ref{smallx}). In order to obtain the intensity-spectra $x(\d N/\d x)$, we still need to perform the integrals over $\d^2\vc k_\perp$, $\d^2\vc q_\perp$ and $\d\Delta z$. The final-state splitting kernels presented in~\cite{Ovanesyan:2011kn} allowed an analytical integration using certain assumptions. Due to the complication with the effective masses this is not possible for the initial-state splitting kernels. Instead, we will perform the remaining integrations numerically in the next Section~\ref{sec:numerics}. Finally, we note that unlike the final-state splitting kernels which are finite in the limit $m_{\text{eff}}=0$, this is not the case for the initial-state splitting kernels that are logarithmically divergent in the mass. This is true both in the finite-$x$ and SGA limit~\cite{Vitev:2007ve}.

\section{Numerical Results} \label{sec:numerics}
\begin{figure*}[!t]
\includegraphics[width= 220pt]{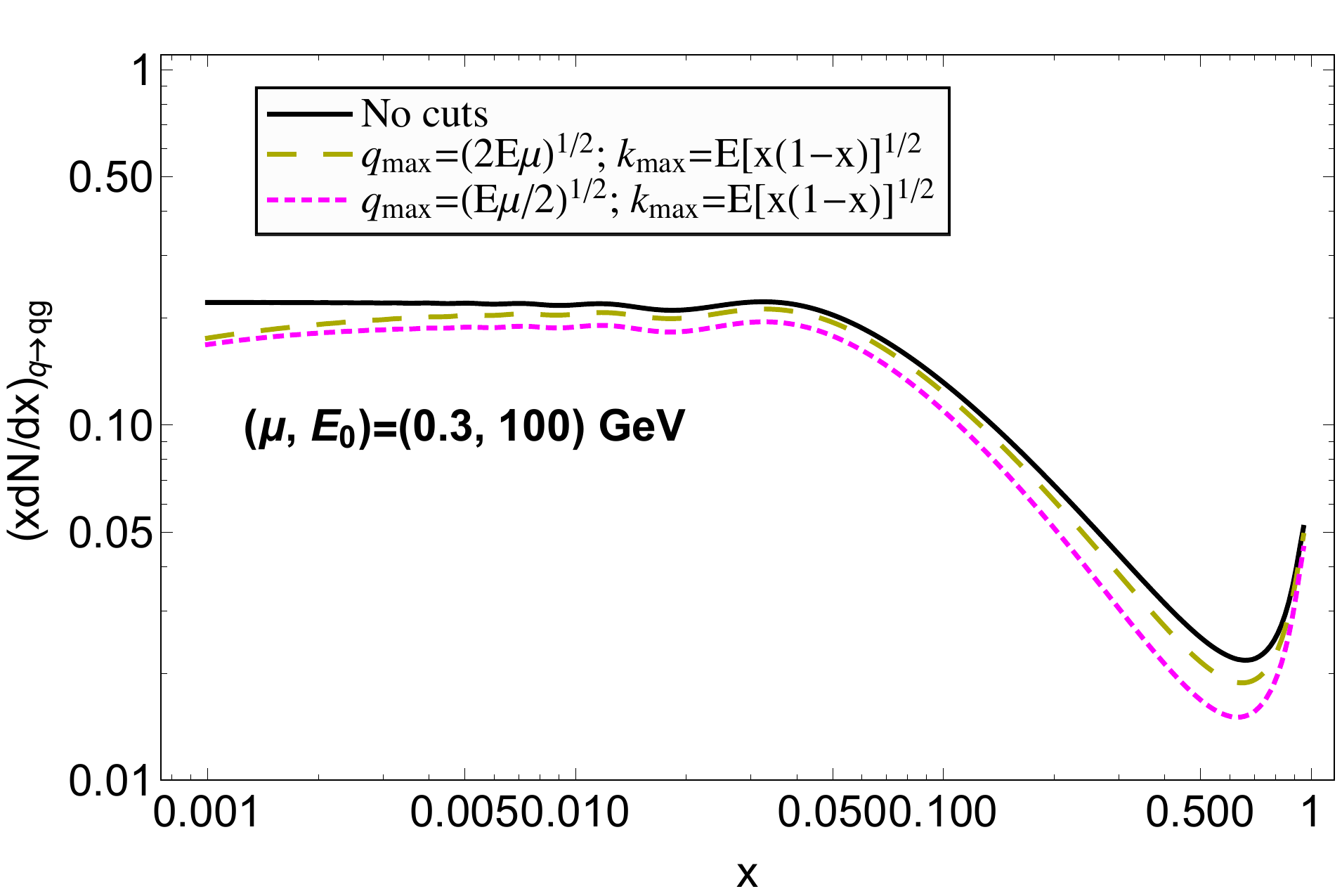} \hspace*{0.1in}\includegraphics[width= 220pt]{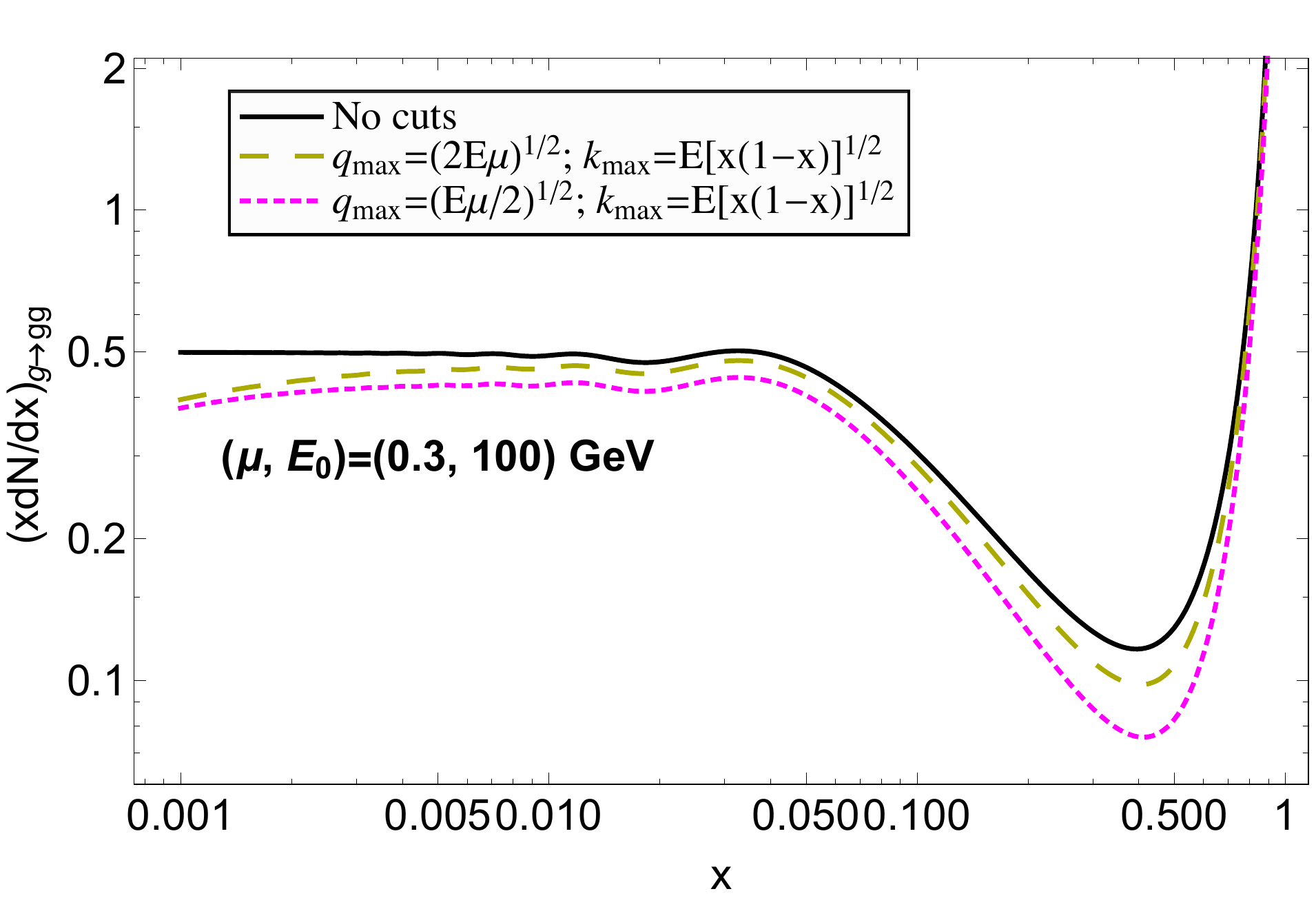}\\
\includegraphics[width= 220pt]{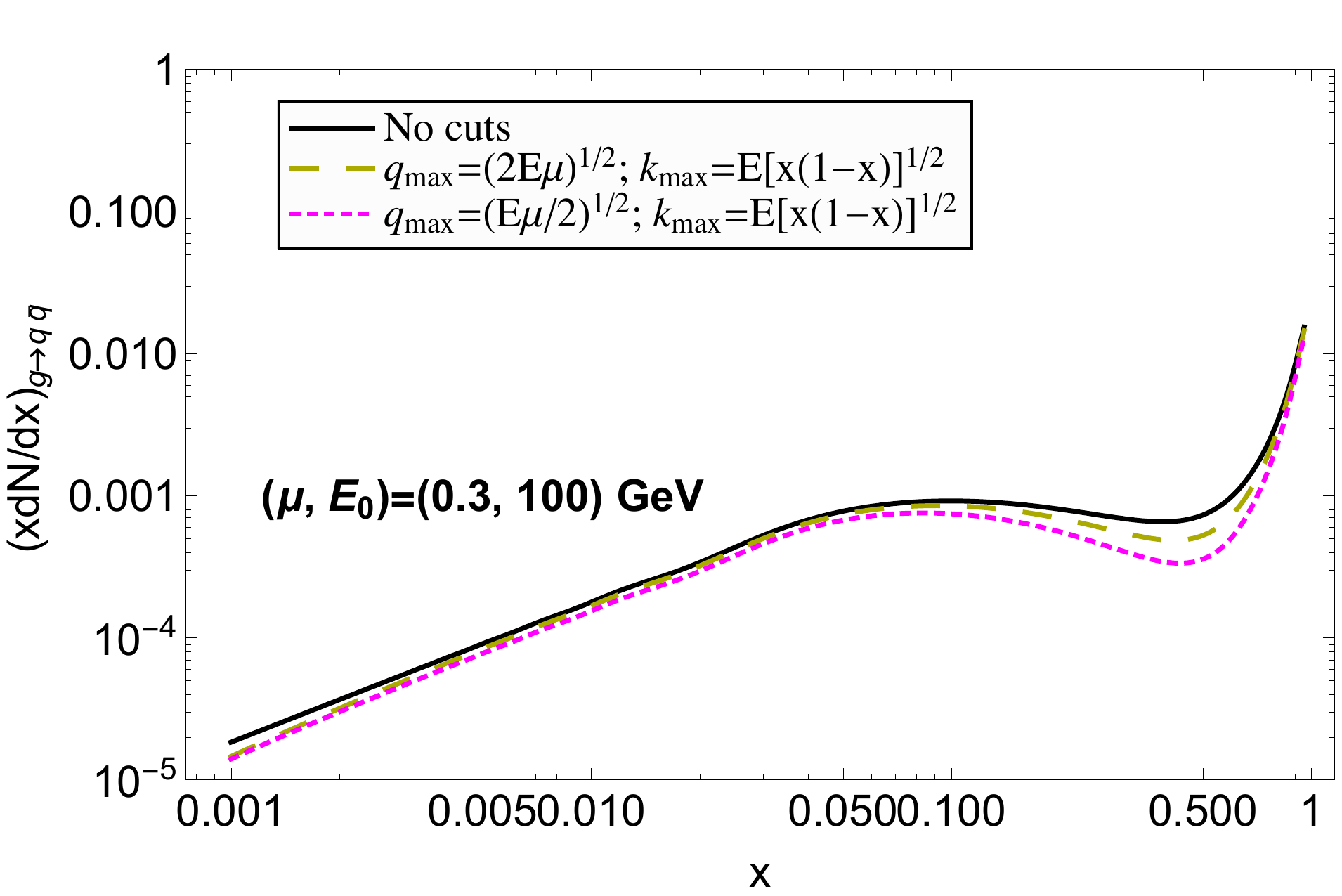} \hspace*{0.1in} \includegraphics[width= 220pt]{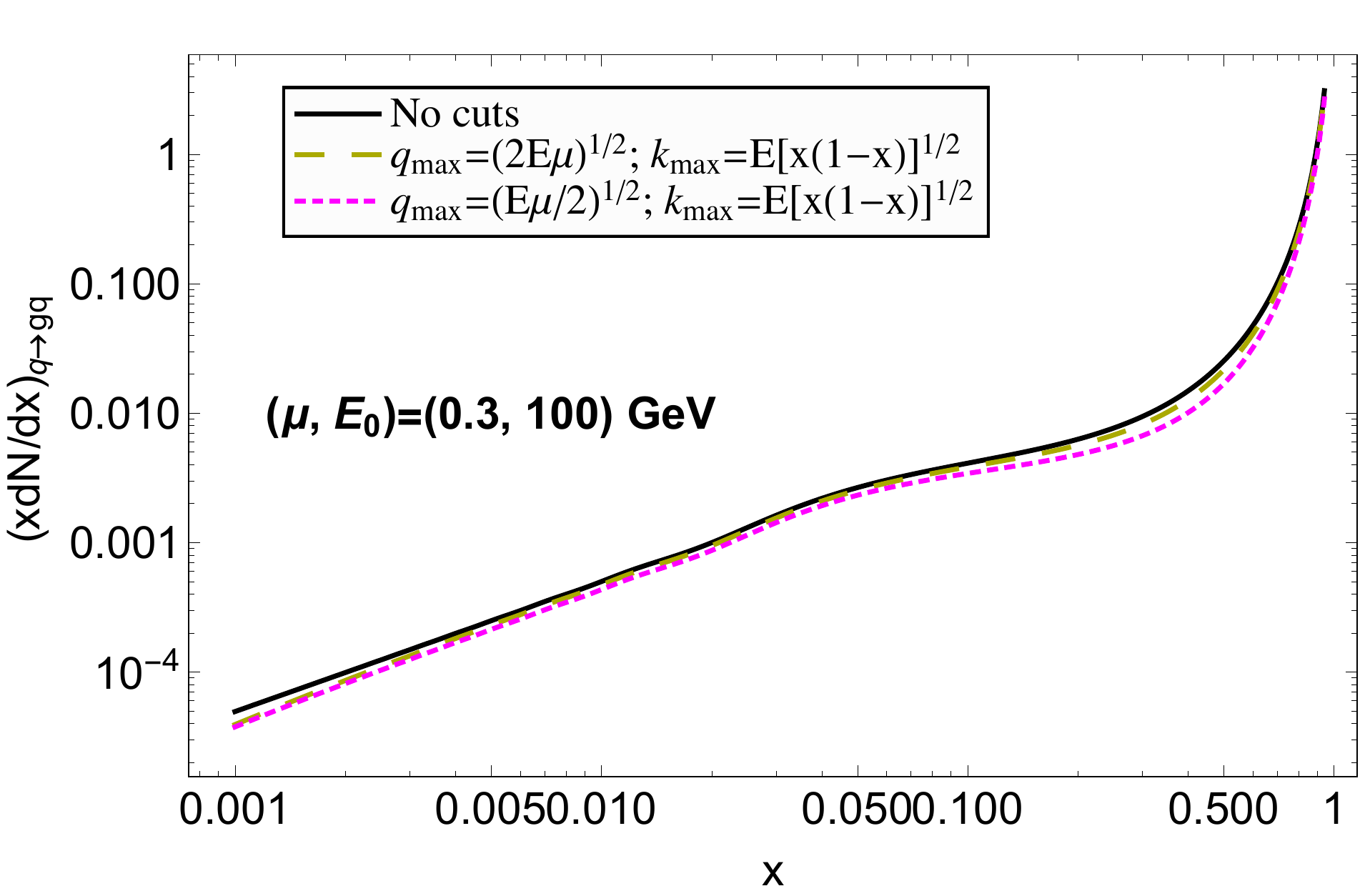}
\caption{Comparison of the intensity spectra $x(\d N/\d x)$ with (olive, magenta) and without (solid black) physical phase space cuts for all four full splitting kernels. We show the results for $k_{\mathrm{max}}=E_0\sqrt{x(1-x)}$, whereas two possible choices of cuts for the momentum $q$ are shown $q_{\mathrm{max}}=\sqrt{2E_0\mu}$ (dashed olive) and $q_{\mathrm{max}}=\sqrt{E_0\mu/2}$ (dotted magenta). We show the results for an energy of $E_0=100$~GeV and a momentum transfer of $\mu=0.3$~GeV.}
\label{fig:numerics2}
\end{figure*}

\begin{figure*}[!t]
\includegraphics[width= 220pt]{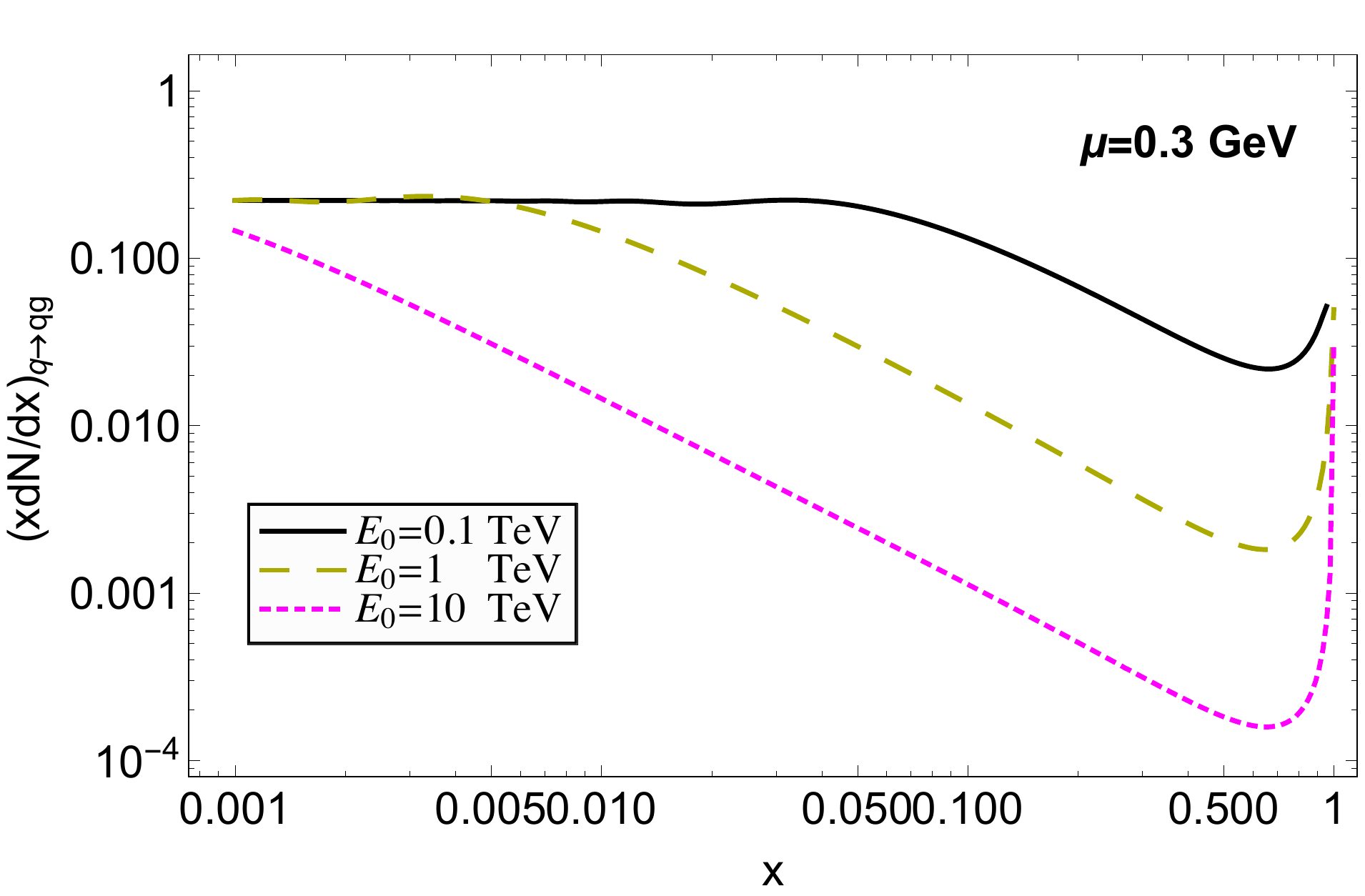} \hspace*{0.1in}\includegraphics[width= 220pt]{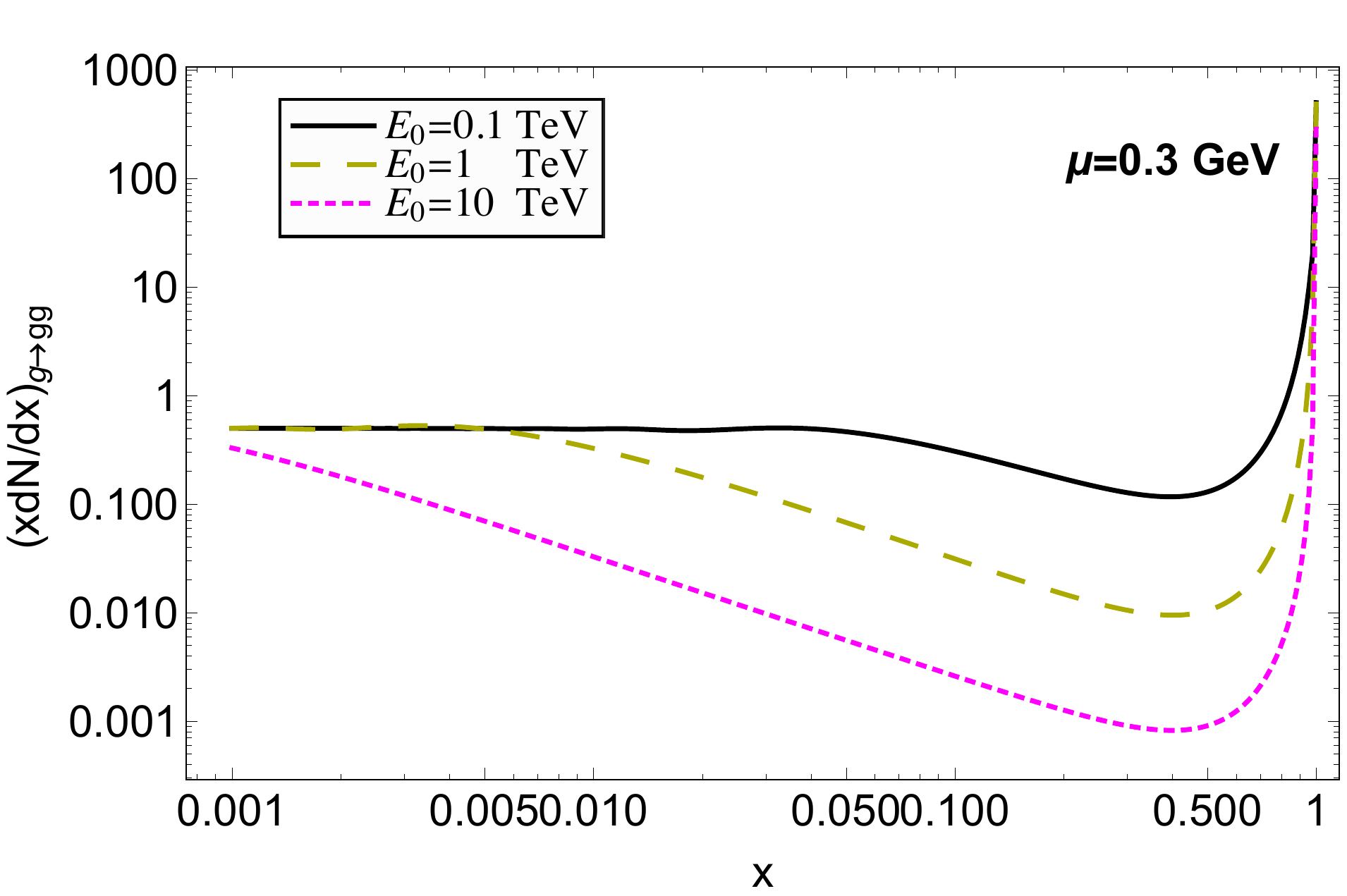}\\
\includegraphics[width= 220pt]{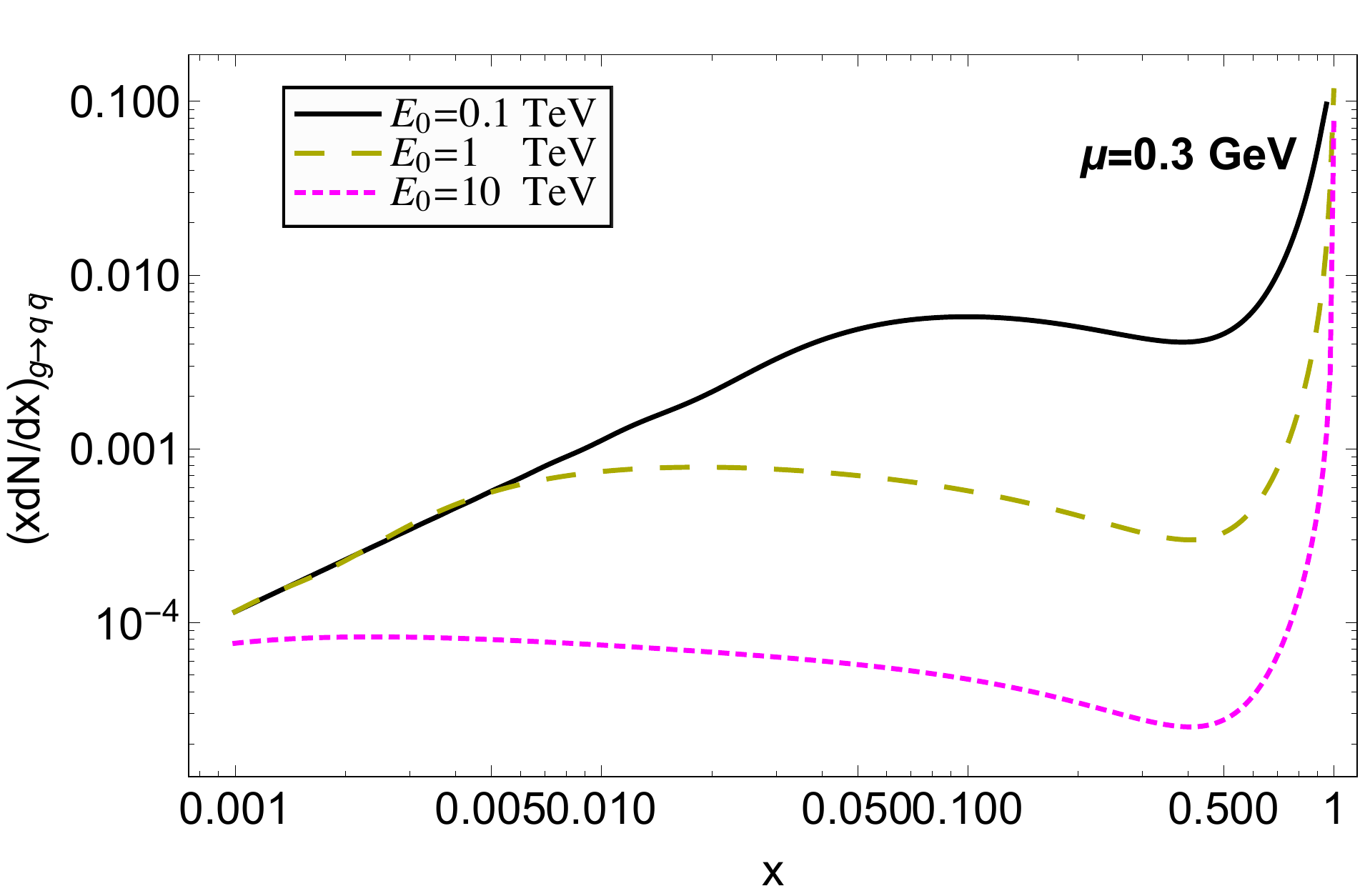} \hspace*{0.1in} \includegraphics[width= 220pt]{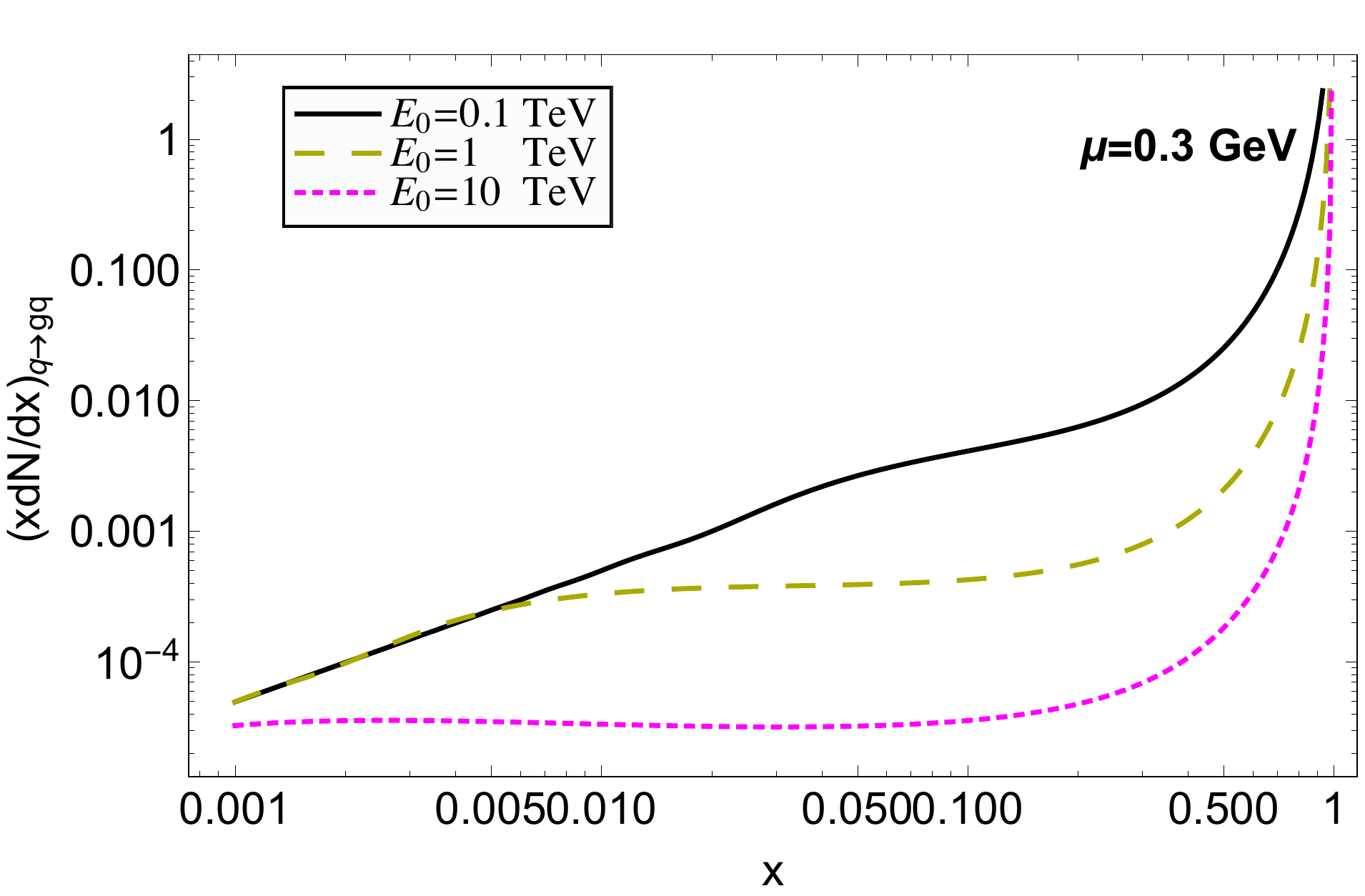}
\caption{The intensity spectra $x(\d N/\d x)$ for all four full splitting kernels for three different energies of the parent parton: $E_0=0.1$~TeV (solid black), $E_0=1$~TeV (dashed olive), $E_0=10$~TeV (dotted magenta).}
\label{fig:numerics3}
\end{figure*}
In this Section, we analyze numerically the effect of our newly derived initial-state splitting kernels at finite-$x$ compared to the small-$x$ kernels used in the traditional energy loss approach. We start with the assumption of no phase space cuts, which we relax afterwards. In addition, we also analyze the energy dependence of the intensity spectra $x(\d N/\d x)$. We also would like to stress that our numerical results are obtained without imposing any positivity constraints on the splitting kernels $x(\d N/\d x/\d^2\vc k_\perp)\geq 0$ as it was done for the small-$x$ results in~\cite{Vitev:2007ve}.

Throughout this Section, we assume that the scattering length $\lambda_{q,g}$ for quarks and gluons is independent of the position $\Delta z$. This assumption corresponds to a uniform static QCD medium. For definiteness, we choose the following numerical values for the medium parameters: $\alpha_s=0.3$ for the strong coupling constant, $\lambda_g=1$~fm for the gluon mean free path, $m_{\text{eff}}=0.94$~GeV for the parton mass acquired in the medium and $L=5$~fm for the size of the medium.

We start by considering no additional phase space cuts for the $\d^2\vc k_\perp$, $\d^2\vc q_\perp$ integrals, hence, we perform the integration up to infinity. In addition, we neglect recoil effects in the medium which allows us to write the elastic scattering cross section as
\be
\frac{1}{\sigma_{\text{el}}}\frac{\d\sigma_{\text{el}}}{\d^2{\vc q}_{\perp}} = \f{\mu^2}{\pi(\vc q_\perp^2+\mu^2)^2}\, ,
\ee
where $\mu$ is the momentum transfer. In Fig.~\ref{fig:numerics1}, we present numerical results for the intensity spectra $x(\d N/\d x)$ for all four splitting kernels. We choose a parent parton energy of $E_0=p_0^+/2=100$~GeV and we plot the results for three different values of the momentum transfer $\mu=0.1,\, 0.3,\, 0.5$~GeV (red, green, blue). We show both the finite-$x$ results (solid) as well as the small-$x$ results (SGA, dashed). Note that the intensity spectra for the diagonal splitting kernels remain flat for small-$x$, whereas the off-diagonal ones fall off linearly on the logarithmic scale used for the plots in Fig.~\ref{fig:numerics1}. As expected, we find that our new results differ the most from the small-$x$ approximations at large values of $x$. The difference between the finite- and small-$x$ intensity spectra changes sign for both processes $q\to qg$ and $g\to q\bar q$. One clearly notices the sensitivity on the momentum transfer $\mu$ of the results for both finite- and small-$x$. The general size of the finite-$x$ corrections are comparable to those found in~\cite{Ovanesyan:2011kn} for the final state. One notices soft oscillations of the intensity spectra which are indeed physical and not a numerical artifact. They can be traced back to the cosine functions that appear in Eqs.~(\ref{eq:qqg})-(\ref{eq:ggggqq}) and~(\ref{smallx}).

In order to check the validity of our numerical integration for the inclusive spitting kernels we performed the following consistency check for the case of no cuts. Firstly, we performed all five integrations, $\d^2\vc k_\perp$, $\d^2\vc q_\perp$ and $\d\Delta z$, numerically and secondly, we performed four out of the five integrations analytically and only one numerically. For all four splitting kernels we found perfect agreement between these two methods for the entire range $x\in(0,1)$. 

In Fig.~\ref{fig:numerics2}, we fix the energy and the momentum transfer to $E_0=100$~GeV and $\mu=0.3$~GeV respectively. In these plots, we study the effects of applying physical phase space cuts instead of the infinite phase space cuts as employed in Fig.~\ref{fig:numerics1}. The results for the intensity spectra $x(\d N/\d x)$ for all four full splitting kernels are shown. We use a generic upper limit $k_{\mathrm{max}}=E_0\sqrt{x(1-x)}$ and for the momentum $q$, we show the results for two different choices $q_{\mathrm{max}}=\sqrt{2E_0\mu}$ (dashed olive) and $q_{\mathrm{max}}=\sqrt{E_0\mu/2}$ (dotted magenta). The reference results for infinite phase space cuts are shown in solid black. We find sizable differences of the same order for all four splitting processes. The effects for the off-diagonal ones $q\to gq$ and $g\to q\bar q$ seem less pronounced only due to the larger scale span on the vertical axis. In addition, we note that finite phase space cuts affect the results of the intensity spectra for all values $x$. This is similar to the final-state splitting intensities in~\cite{Ovanesyan:2011kn}. 

Finally, we show the energy dependence of the intensity spectra in Fig.~\ref{fig:numerics3}. We use again infinite phase space cuts as in Fig.~\ref{fig:numerics1} and we fix the momentum transfer as $\mu=0.3$~GeV. For all four full splitting kernels, we show three different results for the energies of the parent parton $E_0=0.1$~TeV (solid black), $E_0=1$~TeV (dashed olive), $E_0=10$~TeV (dotted magenta). From these figures, we conclude that the probability for flavor changing processes is significant at all energies. The effect of energy redistribution within an initial-state medium induced parton shower of moderate and small momentum fractions $x$ would be most relevant for experiments at relatively low energies, such as fixed target experiments as well as collider experiments at RHIC. At very high energies Bertsch-Gunion type bremsstrahlung from beam jets~\cite{Gyulassy:2014cfa,Gunion:1981qs} is expected to play a dominant role. 

\section{Conclusions} \label{sec:conclusions}

There is significant renewed interest in understanding cold nuclear matter effects in reactions with nuclei 
over a wide range of fixed target and collider energies. So far, soft inelastic parton scattering has been 
primarily treated in the traditional energy loss approach. Here, we extended the calculation 
of parton splitting in cold nuclear matter beyond the soft gluon approximation using the effective theory \SCETG and presented finite-$x$ initial-state splitting kernels to first order in opacity. In addition, we studied the numerical impact of our newly derived full splitting kernels and found sizable differences to previous results in the literature based on the soft-gluon approximation. We also found that employing the correct physical phase space cuts has an important effect for the intensity spectra $x(\d N/\d x)$ for all values of $x$. With the initial-state splitting functions at hand, a fully consistent extension of the traditional approach to energy loss in the QCD medium can be achieved using DGLAP evolution techniques. An analysis of the effect of the full splitting functions for physical observables will be left for future work. We plan to study several phenomenological applications for observables in both p+A and A+A collisions. In addition, we will extend the calculations presented in this work to heavy quarks for both the initial and final state.

%%%%%%%%%%%%%%%%
\medskip
%%%%%%%%%%%%%%%%%%%%%%%%%%%%%%

{\bf Acknowledgments}

This research is supported by the US Department of Energy, Office of Science under Contract No. DE-AC52- 06NA25396 and by the DOE Early Career Program under Grant No. 2012LANL7033.

\bibliographystyle{h-physrev5}
\bibliography{bibliography}

\end{document}